# Confocal structured illumination microscopy

Weishuai Zhou,[1] Manhong Yao,[2] Xi Lin,[1] Quan Yu,[3] Junzheng Peng,[1,*] and Jingang Zhong,[1,*]

[1] Department of Optoelectronic Engineering, Jinan University, Guangzhou 510632, China

[2] School of Optoelectronic Engineering, Guangdong Polytechnic Normal University, Guangzhou 510665, China

[3] School of Medicine, Jinan University, Guangzhou 510632, China

*Authors to whom correspondence should be addressed: junzpeng@jnu.edu.cn and tzjg@jnu.edu.cn

**Abstract:** Confocal microscopy, a critical advancement in optical imaging, is widely applied because of its excellent anti-noise ability. However, it has low imaging efficiency and can cause phototoxicity. Optical-sectioning structured illumination microscopy (OS-SIM) can overcome the limitations of confocal microscopy but still face challenges in imaging depth and signal-to-noise ratio (SNR). We introduce the concept of confocal imaging into OS-SIM and propose confocal structured illumination microscopy (CSIM) to enhance the imaging performance of OS-SIM. CSIM exploits the principle of dual photography to reconstruct a dual image from each pixel of the camera. The reconstructed dual image is equivalent to the image obtained by using the spatial light modulator (SLM) as a virtual camera, enabling the separation of the conjugate and non-conjugate signals recorded by the camera pixel. We can reject the non-conjugate signals by extracting the conjugate signal from each dual image to reconstruct a confocal image when establishing the conjugate relationship between the camera and the SLM. We have constructed the theoretical framework of CSIM. Optical-sectioning experimental results demonstrate that CSIM can reconstruct images with superior SNR and greater imaging depth compared with existing OS-SIM. CSIM is expected to expand the application scope of OS-SIM.

Since Marvin Minsky first proposed confocal imaging in 1957 [1], this technique has evolved to become an indispensable tool in disciplines requiring high-precision optical microscopy [2,3] and is a milestone in traditional spatial domain imaging. For fluorescent biological imaging, the laser scanning confocal microscope (LSCM) offers three-dimensional (3D) imaging with improved signal-to-noise ratio and reduced background noise. Despite these benefits, LSCM's spot scanning strategy leads to low imaging efficiency and phototoxicity concerns [3].

In the past two decades, various spatial-frequency domain imaging techniques based on structured illumination have emerged [4–6]. Spatial-frequency domain imaging is promising to overcome the limitations inherent in traditional spatial domain imaging. An exemplary breakthrough in this field is Fourier single-pixel imaging [7]. It reconstructs the image of the sample by using Fourier basis patterns for structured illumination, obtaining the Fourier spectrum of the sample image in the spatial-frequency domain, and performing an inverse Fourier transform on the acquired Fourier spectrum. Fourier single-pixel imaging can be used to overcome the limitation induced by the pixel array detector [8–10]. Other imaging techniques based on structured illumination can also be broadly categorized as spatial-frequency domain imaging techniques, such as optical-sectioning structured illumination microscopy (OS-SIM) [4,11–13], super-resolution structured illumination microscopy (SR-SIM) [5,6,14–16], structured illumination profilometry [17–19], and structured illumination imaging used for determining the optical properties of biological tissues [20,21]. Nevertheless, the spatial-frequency domain imaging has inherent limitations. Many techniques have been developed to overcome these limitations, such as space-frequency jointed algorithms [22,23]. However, OS-SIM still suffers from shallow imaging depth and low signal-to-noise ratio of imaging results [24,25].

We introduce the concept of confocal imaging into spatial-frequency domain imaging and propose confocal structured illumination microscopy (CSIM) to enhance the imaging performance of OS-SIM. Although the experimental setup of CSIM is the same as that of the OS-SIM, CSIM has a different imaging mechanism and image reconstruction algorithm. CSIM exploits the principle of dual photography [26] to reconstruct a dual image from each pixel of the camera by Fourier single-pixel imaging. The reconstructed dual image is equivalent to the image captured by using SLM as a virtual camera, enabling the separation of the non-conjugate and conjugate signals recorded by the camera pixel. We can reject non-conjugate signals by extracting the conjugate signals from the dual images once we establish the conjugate relationship between the camera and the SLM. By rearranging the extracted conjugate signals according to the camera pixel coordinates, we can reconstruct a confocal image. Both theoretical analysis and experimental validation demonstrate that CSIM achieves a higher signal-to-noise ratio (SNR) result and greater imaging depth compared with existing OS-SIM. The proposed CSIM paves the way for advancing structured illumination microscopy.

We use Fig. 1 to illustrate the imaging principle of CSIM. The light beam from the light source (LS) illuminates the sample (S) through an illumination optical system consisting of a collimating lens (CL), an SLM, an illuminating tube lens (ITL), and an illuminating objective lens (IOL). The sample (S) is imaged onto the sensor plane of a camera (C) by a detection optical system consisting of a detection objective lens (DOL) and a detecting tube lens (DTL). The illumination optical system and the detection optical system are coaxial. The object plane, the modulation plane of the SLM, and the sensor plane of the camera are optically conjugate to each other. Thus, the object points on the object plane (e.g., A and B) are conjugating with the pixels on the sensor plane of the camera (e.g., A″ and B″) and are also conjugating with the pixels on the modulation plane of the SLM (e.g., A′ and B′).

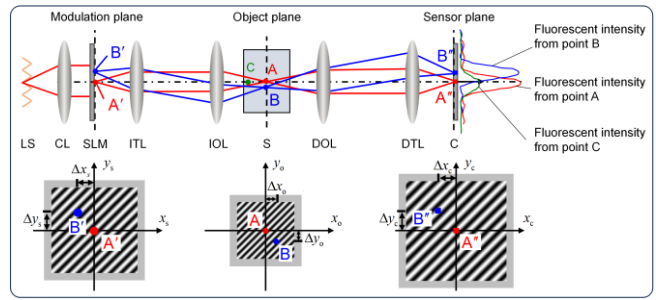

Fig. 1 Principle of CSIM. (LS: light source; CL: collimating lens; ITL: illuminating tube lens; IOL: illuminating objective lens; S: sample; DOL: detecting objective lens; DTL: detecting tube lens; C: camera.)

To reconstruct the dual image by using Fourier single-pixel imaging, we load a set of Fourier basis patterns on the modulation plane of the SLM to generate Fourier-structured light. The Fourier basis pattern can be expressed as:

$$P(x_s, y_s; f_{x_s}, f_{y_s}; \phi_k) = a + b\cos(2\pi f_{x_s} x_s + 2\pi f_{y_s} y_s + \varphi_0 + \phi_k), \tag{1}$$

where $(x_s, y_s)$ represents the spatial coordinate on the modulation plane of the SLM, $(f_{x_s}, f_{y_s})$ represents the spatial frequency. $a$ represents the background intensity level, $b$ represents contrast, $\varphi_0$ represents the initial phase, and $\phi_k = 2\pi k/4$ represents the phase shift, $k = 0, 1, 2, 3$.

The intensity distribution of the Fourier-structured light at the object plane can be represented as follows:

$$I_o(x_o, y_o; f_{x_s}, f_{y_s}; \phi_k) = \iint P(x_s, y_s; f_{x_s}, f_{y_s}; \phi_k) h_{ill}\left(x_o - \frac{x_s}{\alpha}, y_o - \frac{y_s}{\alpha}\right) dx_s dy_s, \tag{2}$$

where $(x_o, y_o)$ represents the spatial coordinate on the object plane. $h_{ill}(x_o, y_o)$ and $\alpha$ represent the intensity point spread function and the magnification of the illumination optical system, respectively.

The intensity value recorded by each pixel of the camera can be expressed as follows:

$$I(x_c, y_c; f_{x_s}, f_{y_s}; \phi_k) = \eta \iint \left\{ \iint \begin{bmatrix} P(x_s, y_s; f_{x_s}, f_{y_s}; \phi_k) \\ h_{ill}\left(x_o - \frac{x_s}{\alpha}, y_o - \frac{y_s}{\alpha}\right) \end{bmatrix} dx_s dy_s \times O(x_o, y_o) h_{dect}\begin{pmatrix} x_c - \beta x_o, \\ y_c - \beta y_o \end{pmatrix} \right\} dx_o dy_o, \tag{3}$$

where $(x_c, y_c)$ represents the coordinate of pixel on the sensor plane of the camera, $h_{dect}(x_c, y_c)$ and $\beta$ represent the intensity point spread function and the magnification of the detection optical system, respectively. $O(x_o, y_o)$ represents the object function. $\eta$ is a factor that depends on the size and the location of the detector.

If we generate four Fourier basis patterns with spatial frequency $(f_{x_s}, f_{y_s})$ and phases $\phi_k$ of $0$, $\pi/2$, $\pi$, $3\pi/2$ in turn, we can record four intensity values by each pixel of the camera. With the four intensity values, we can calculate the Fourier coefficient at the spatial frequency $(f_{x_s}, f_{y_s})$:

$$\begin{aligned} &\left[I(x_c, y_c; f_{x_s}, f_{y_s}; 0) - I(x_c, y_c; f_{x_s}, f_{y_s}; \pi)\right] \\ &+ j\left[I(x_c, y_c; f_{x_s}, f_{y_s}; \pi/2) - I(x_c, y_c; f_{x_s}, f_{y_s}; 3\pi/2)\right] \\ &= 2b\eta \iint \left\{ \iint \begin{bmatrix} h_{ill}\left(x_o - \frac{x_s}{\alpha}, y_o - \frac{y_s}{\alpha}\right) O(x_o, y_o) \\ \times h_{dect}(x_c - \beta x_o, y_c - \beta y_o) \end{bmatrix} dx_o dy_o \right\} dx_s dy_s \\ &\times e^{-j\varphi_0} e^{-j2\pi(f_{x_s} x_s + f_{y_s} y_s)} \\ &= \tilde{I}_{(x_c, y_c)}(f_{x_s}, f_{y_s}) \end{aligned} \tag{4}$$

Equation (4) is the four-step phase-shift algorithm. We can also use the three-step phase-shifting algorithm [27] to calculate the Fourier coefficients.

By changing the spatial frequency of the Fourier basis pattern, we can obtain all the Fourier coefficients. However, since the fluorescence image is real-valued, the spectrum exhibits conjugate symmetry. Therefore, we only need to obtain half of the Fourier coefficients. The remaining coefficients can be obtained by using conjugate symmetry. Based on the acquired Fourier coefficients, we can construct a Fourier spectrum. Applying the inverse Fourier transform to the acquired Fourier spectrum, we can reconstruct an image:

$$I_{(x_c,y_c)}(x_s,y_s) = 2b\eta\cos(\varphi_0)\iint h_{\text{ill}}\left[\begin{pmatrix}x_o - \frac{x_s}{\alpha}, \\ y_o - \frac{y_s}{\alpha}\end{pmatrix} O(x_o,y_o) h_{\text{det}}\begin{pmatrix}x_c - \beta x_o, \\ y_c - \beta y_o\end{pmatrix}\right]dx_o dy_o, \qquad (5)$$

$$= \text{Real}\left\{\text{IFT}\left[\tilde{I}_{(x_c,y_c)}(f_{x_s},f_{y_s})\right]\right\}$$

where $\text{IFT}[\cdot]$ represents the 2D inverse Fourier transform operator, $\text{Real}\{\cdot\}$ represents the take real part operator.

The reconstructed image $I_{(x_c,y_c)}(x_s,y_s)$ is a dual image according to the principle of dual photography [26]. If the camera has $M_c \times N_c$ pixels, we can reconstruct $M_c \times N_c$ dual images.

The reconstructed dual image $I_{(x_c,y_c)}(x_s,y_s)$ can separate the signals of the non-conjugate object point and the signal of the conjugate object point corresponding to the camera pixel $(x_c,y_c)$. For example, using the camera pixel $A''$ shown in Fig. 1, the reconstructed dual image $I_{A''}(x_s,y_s)$ can separate the signal from the conjugate object point A and the signal from the non-conjugate object point B, because the spatial distribution of the reconstructed dual image is determined by the structured light, and the object points A and B are modulated by the structured light generated by the pixels $A'$ and $B'$ on the SLM, respectively.

In addition, most of the signals from the out-of-focus object point, such as C in Fig. 1, do not appear in the dual image $I_{(x_c,y_c)}(x_s,y_s)$, because the camera's pixel size is only a few microns, with each pixel of the camera acting as a detection pinhole in confocal imaging.

We can reject the non-conjugate signals by extracting the conjugate signals from dual images to construct a confocal image once we establish the conjugate relationship between the SLM and camera:

$$I_{\text{conf}}(x_c,y_c) = I_{(x_c,y_c)}(x_s,y_s)\delta(x_s - x_{s\text{-c}}, y_s - y_{s\text{-c}}),$$
$$= I_{(x_c,y_c)}(x_{s\text{-c}},y_{s\text{-c}}) \qquad (6)$$

where $(x_{s\text{-c}},y_{s\text{-c}})$ represents the spatial coordinate of the pixel on the SLM conjugated to the camera pixel $(x_c,y_c)$.

On the other hand, the reconstructed dual image by Eq. (5) can be rewritten as follows:

$$I_{(x_c,y_c)}(x_s,y_s)$$
$$= \text{Real}\left\{\text{IFT}\left[\tilde{I}_{(x_c,y_c)}(f_{x_s},f_{y_s})\right]\right\}$$
$$= \text{Real}\left\{\iint \tilde{I}_{(x_c,y_c)}(f_{x_s},f_{y_s})e^{j2\pi(f_{x_s}x_s+f_{y_s}y_s)}df_{x_s}df_{y_s}\right\}, \qquad (7)$$
$$= \iint \cos\left\{\begin{array}{l}2\pi(f_{x_s}x_s + f_{y_s}y_s) + \\ \text{angle}\left[\tilde{I}_{(x_c,y_c)}(f_{x_s},f_{y_s})\right]\end{array}\right\}\left|\tilde{I}_{(x_c,y_c)}(f_{x_s},f_{y_s})\right|df_{x_s}df_{y_s}$$

where $\left|\tilde{I}_{(x_c,y_c)}(f_{x_s},f_{y_s})\right|$ and $\text{angle}\left[\tilde{I}_{(x_c,y_c)}(f_{x_s},f_{y_s})\right]$ represent the modulus and angle of the Fourier coefficient $\tilde{I}_{(x_c,y_c)}(f_{x_s},f_{y_s})$, respectively.

Combining Eqs. (6) and (7), we can rewrite the confocal image as follows:

$$I_{\text{conf}}(x_c,y_c) = \sum_{f_{x_s}=-\infty}^{f_{x_s}=+\infty}\sum_{f_{y_s}=-\infty}^{f_{y_s}=+\infty}\cos\left\{\begin{array}{l}2\pi(f_{x_s}x_{s\text{-c}} + f_{y_s}y_{s\text{-c}}) + \\ \text{angle}\left[\tilde{I}_{(x_c,y_c)}(f_{x_s},f_{y_s})\right]\end{array}\right\}\left|\tilde{I}_{(x_c,y_c)}(f_{x_s},f_{y_s})\right|. \qquad (8)$$

Given that natural image spectra exhibit conjugate symmetry,

$$\tilde{I}_{(x_c,y_c)}(f_{x_s}, f_{y_s}) = \tilde{I}_{(x_c,y_c)}(-f_{x_s}, -f_{y_s}), \tag{9}$$

we can simplify Eq. (8) as follows:

$$I_{\text{conf}}(x_c, y_c) = \sum_{f_{x_s}=0}^{f_{x_s}=+\infty} \sum_{f_{y_s}=0}^{f_{y_s}=+\infty} 2\cos\left\{\begin{array}{l} 2\pi(f_{x_s}x_{s\text{-c}} + f_{y_s}y_{s\text{-c}}) + \\ \text{angle}\left[\tilde{I}_{(x_c,y_c)}(f_{x_s}, f_{y_s})\right] \end{array}\right\} \left|\tilde{I}_{(x_c,y_c)}(f_{x_s}, f_{y_s})\right|. \tag{10}$$

Equation (10) suggests that we do not need to reconstruct a dual image from each camera pixel for confocal imaging. Instead, we can directly reconstruct a confocal image by substituting the calculated Fourier coefficients and the coordinates of conjugate points obtained by calibration into Eq. (10). This substantially reduces the image reconstruction time of CSIM.

The imaging efficiency of CSIM can be further improved by using a small amount of structured light with different frequencies or even a single-frequency structured light. Let's first deduce the result obtained by using a single-frequency structured light. The illumination optical system and detection optical system of the CSIM are coaxial (Fig. 1). When we disregard diffraction effects and reference Eq. (3), we can rewrite the intensity recorded by the camera pixel $(x_c, y_c)$ as follows:

$$I(x_c, y_c; f_{x_s}, f_{y_s}; \phi_k) = \eta O(x_c, y_c) \left[\begin{array}{l} a + b\cos(2\pi f_{x_s} x_{s\text{-c}} \\ + 2\pi f_{y_s} y_{s\text{-c}} + \varphi_0 + \phi_k) \end{array}\right], \tag{11}$$

where $O(x_c, y_c)$ represents the image of the sample on the sensor plane of the camera. Combing Eqs. (4) and (11), we can yield the following expression:

$$\begin{array}{l} \left[I(x_c, y_c; f_{x_s}, f_{y_s}; 0) - I(x_c, y_c; f_{x_s}, f_{y_s}; \pi)\right] \\ + j\left[I(x_c, y_c; f_{x_s}, f_{y_s}; \pi/2) - I(x_c, y_c; f_{x_s}, f_{y_s}; 3\pi/2)\right] \\ = 2b\eta O(x_c, y_c) e^{-j\varphi_0} e^{-j2\pi(f_{x_s}x_{s\text{-c}} + f_{y_s}y_{s\text{-c}})} \\ = \tilde{I}_{(x_c,y_c)}(f_{x_s}, f_{y_s}) \end{array} \tag{12}$$

From Eq. (12), we can conclude that:

$$\text{angle}\left[\tilde{I}_{(x_c,y_c)}(f_{x_s}, f_{y_s})\right] = -\varphi_0 - 2\pi(f_{x_s}x_{s\text{-c}} + f_{y_s}y_{s\text{-c}}), \tag{13}$$

$$\left|\tilde{I}_{(x_c,y_c)}(f_{x_s}, f_{y_s})\right| = 2b\eta O(x_c, y_c). \tag{14}$$

And consequently, we have

$$\cos\left\{\begin{array}{l} 2\pi(f_{x_s}x_{s\text{-c}} + f_{y_s}y_{s\text{-c}}) + \\ \text{angle}\left[\tilde{I}_{(x_c,y_c)}(f_{x_s}, f_{y_s})\right] \end{array}\right\} \left|\tilde{I}_{(x_c,y_c)}(f_{x_s}, f_{y_s})\right| = 2b\eta \cos(\varphi_0) O(x_c, y_c). \tag{15}$$

Equation (15) suggests that the confocal image of the sample can be reconstructed using CSIM with a single-frequency structured light. However, similar to OS-SIM, using a single-frequency structured light in CSIM may result in low SNR results for thick or densely labeled samples [28]. To achieve high SNR results, we can use multiple structured lights with different frequencies. These spatial frequencies are arranged in a ring in the Fourier domain. Increasing the radius of the ring results in a higher spatial frequency. To enhance the resolution of the reconstructed image, we can use a larger radius for the ring. By following this approach, we can use the following equation to reconstruct the confocal image:

$$I_{\text{conf}}(x_c, y_c) = \sum_{k=1}^{K} 2\cos\left\{\begin{array}{l} 2\pi(f_{x_s}^k x_{s\text{-c}} + f_{y_s}^k y_{s\text{-c}}) + \\ \text{angle}\left[\tilde{I}_{(x_c,y_c)}(f_{x_s}^k, f_{y_s}^k)\right] \end{array}\right\} \left|\tilde{I}_{(x_c,y_c)}(f_{x_s}^k, f_{y_s}^k)\right|, \tag{16}$$

where $k=1,2,\cdots,K$ is the serial number of the used structured light. $\sqrt{\left(f_{x_s}^k\right)^2 + \left(f_{y_s}^k\right)^2} = R_f$, $R_f$ represents the radius of the ring.

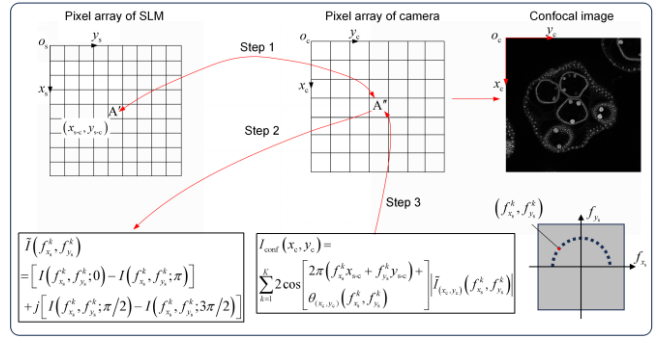

Fig. 2 Image reconstruction process of CSIM.

We use Fig. 2 to summarize the confocal image reconstruction process of CSIM:

Step 1: Calibrate the conjugate relationship between the SLM and the camera.

Step 2: Use Eq. (4) to calculate the Fourier coefficients for each camera pixel.

Step 3: Substitute the coordinates of the conjugated point obtained in Step 1 and the Fourier coefficients obtained in Step 2 into Eq. (16), we can reconstruct the confocal image.

We carried out experiments to verify the proposed CSIM. Figure 3 illustrates the experimental setup, which used an incoherent LED (SOLIS-470C, central wavelength: 470 nm, bandwidth: 34 nm, Thorlabs) and a digital micromirror device (Vialux V6501, $1920 \times 1080$ pixels, pixel size: $7.6\ \mu m$) to generate Fourier-structured light. To minimize the influence of the LED lamp bead image, we placed a ground glass diffuser in front of the LED. The Fourier-structured light passed through an illuminating tube lens (focal length: 400 mm), a dichroic mirror (transmission wavelength: 505 nm-700 nm), and an apochromatic objective lens ($20\times$, numerical aperture: 0.75, Nikon) to excite the sample emitting fluorescence. The fluorescent images were captured by a camera (Andor Neo 5.5, $2560 \times 2160$ pixels, pixel size: $6.5\ \mu m$) following passage through the objective lens, dichroic mirror, emission filter (transmission wavelength range: 510 nm-700 nm), and imaging tube lens (focal length: 200 mm).

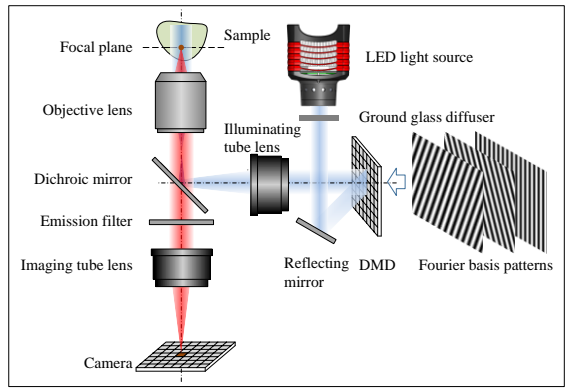

Fig. 3. Experimental setup. (DMD: digital micromirror device.)

To demonstrate the performance of the proposed method, we imaged an autofluorescent sample of volvox with a thickness of about $100\ \mu m$. The spatial periods of the used Fourier basis patterns loaded on the modulation plane of the SLM were from 18.1 to 15.88 pixels, and the corresponding spatial frequencies were from 0.291 to 0.331 $\mu m^{-1}$ on the object plane. We used a three-step phase-shifting algorithm to obtain the Fourier coefficient. To obtain the 3D confocal image of the measured sample, we used a stage (Physik Instrumente (PI), P-737.1SL; Linear travel: 100 μm, Resolution: 2.5 nm, Repeatability: 6 nm) to scan the sample along the axial direction with a scanning interval of 1 μm.

For comparison, wide-field illumination microscopy, LSCM, OS-SIM, and CSIM were employed to reconstruct 3D images of the sample. The image reconstruction equation of OS-SIM can be expressed as follows:

$$I_{\text{OS-SIM}}(x_c, y_c) = \sum_{k=1}^{K} \sqrt{\left[I(x_c, y_c; f_{x_s}^k, f_{y_s}^k; 0) - I(x_c, y_c; f_{x_s}^k, f_{y_s}^k; \pi)\right]^2 + \left[I(x_c, y_c; f_{x_s}^k, f_{y_s}^k; \pi/2) - I(x_c, y_c; f_{x_s}^k, f_{y_s}^k; 3\pi/2)\right]^2}, \quad (17)$$

when $K=1$, Equation (17) represents the classical image reconstruction of OS-SIM [5]. When $K>1$, it means that structured lights of multiple frequencies are used to obtain multiple optical sectioning results and these results are averaged [24].

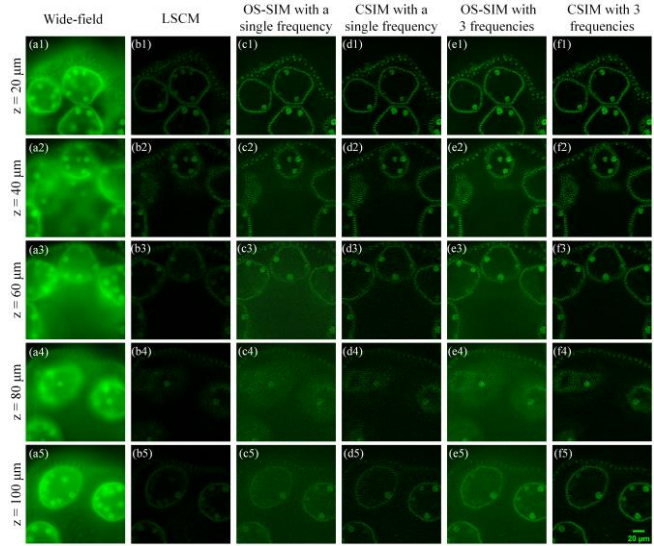

Fig 4. The optical-sectioning results of volvox at depths of 20 μm, 40 μm, 60 μm, 80 μm, and 100 μm using the wide-field illumination microscopy, LSCM, OS-SIM, and CSIM with structured light of single and 3 frequencies.

The LSCM results were obtained by using a commercial confocal microscope equipped with a 470 nm laser and an HCPLAPOCS2 objective lens (magnification: 20, NA: 0.75. The bandwidth of the emission filter was 510 nm to 700 nm, The size of the used pinhole was 1 Airy unit. The size of the reconstructed results was $512 \times 512$ pixels, which cost 262,114 measurements.

Figure 4 shows the experimental results reconstructed by wide-field illumination microscopy, LSCM, OS-SIM, and CSIM with structured light of single and 3 frequencies. The number of measurements required by using structured light with 3 frequencies in OS-SIM and CSIM was 9. The frequencies of the structured light used for the OS-SIM are consistent with that of CSIM. We observe that wide-field illumination microscopy yields images with a strong defocused background. LSCM can suppress defocused background but suffers from a low SNR due to scattering noise. OS-SIM using structured light with a single frequency can reconstruct clear results at 20 μm. However, the SNR of the results decreases quickly with the imaging depth. CSIM using structured light with a single frequency reconstructs a higher SNR compared with OS-SIM. Using structured light with 3 frequencies in OS-SIM smooths the noise but introduces a direct current (DC) background while using structured light with 3 frequencies in CSIM improves the SNR without introducing the DC background.

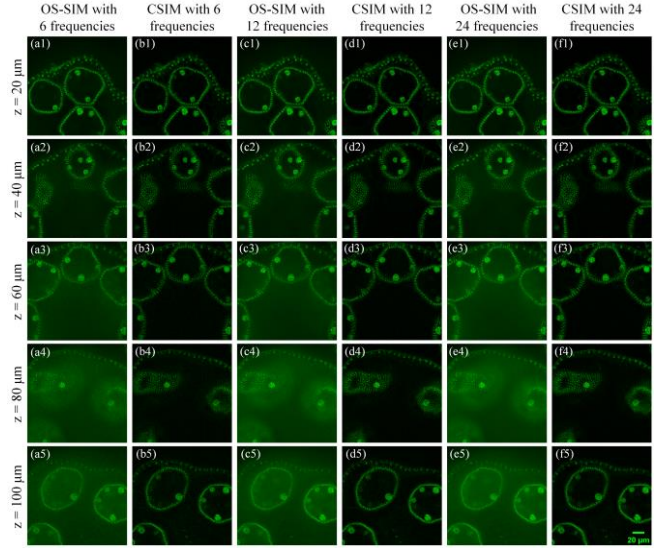

Fig. 5. The optical-sectioning results of volvox at depths of 20 μm, 40 μm, 60 μm, 80 μm, and 100 μm using OS-SIM and CSIM with structured light of 6, 12, and 24 frequencies.

Figure 5 shows the results obtained at different depths by OS-SIM and CSIM using structured light with 6, 12, and 24 frequencies. The numbers of measurements required for these results were 18, 36, and 72, respectively. As seen, increasing frequencies in OS-SIM does not eliminate DC background, while increasing frequencies in CSIM do not induce DC background and reconstruct higher SNR results. In addition, the results obtained by the CSIM with structured light of 6, 12, and 24 frequencies are comparable, suggesting that CSIM using structured light with 6 frequencies is sufficient for high-quality imaging of the Volvox sample.

Figures 6 and 7 show the 3D results of the volvox sample reconstructed using wide-field illumination microscopy, LSCM, OS-SIM, and CSIM using structured light with different frequencies. The 3D result obtained by the wide-field illumination microscopy is hindered by an out-of-focus background. LSCM eliminates out-of-focus background noise but lacks clarity in structure visualization. OS-SIM using structured light with a single frequency provides a clear view of the top structure but suffers from low SNR at the bottom. Conversely, CSIM using structured light with a single frequency allows clear reconstruction of the entire 3D structure of the sample. When OS-SIM uses structured light with multiple frequencies, strong DC background noise masks the sample structure, whereas CSIM using structured light with the same frequencies produces DC background-free results and improved SNR. CSIM using structured light with 6 frequencies yields comparable results to those obtained by CSIM using structured light with 12 and 24 frequencies.

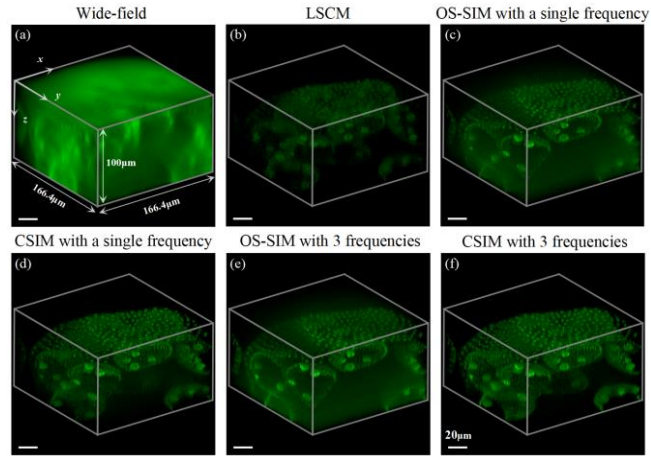

Fig. 6. The 3D results of the volvox sample obtained by wide-field illumination microscopy, LSCM, OS-SIM, and CSIM using structured light with 1, and 3 frequencies, respectively.

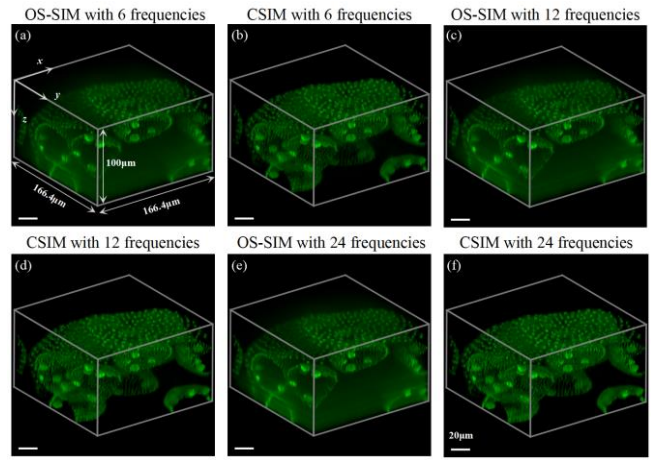

Fig. 7. The 3D results of the volvox sample obtained by OS-SIM and CSIM using structured light with 6, 12, and 24 frequencies, respectively.

To calibrate the resolution of our experimental system, we imaged a fluorescent microsphere sample with a diameter of 200 nm. The spatial frequency of the structured light used was 0.291 μm$^{-1}$. Five fluorescent microspheres were randomly selected from the sample and imaged using wide-field illumination microscopy, OS-SIM, and CSIM. When using OS-SIM and CSIM, a single-frequency structured light and multiple structured lights with 24 frequencies were used, respectively. To obtain 3D results of the microspheres, we used the PI stage to scan the sample axially, with an interval of 0.2 μm and a total of 51 scans. Other experimental parameters were consistent with those used in the previous experiments.

In Fig. 8, the lateral resolution of the results reconstructed by OS-SIM and CSIM are comparable but not improved compared with that of the results reconstructed by wide-field illumination microscopy. The axial resolution of the results reconstructed by OS-SIM and CSIM is higher than that of the result reconstructed by wide-field illumination microscopy.

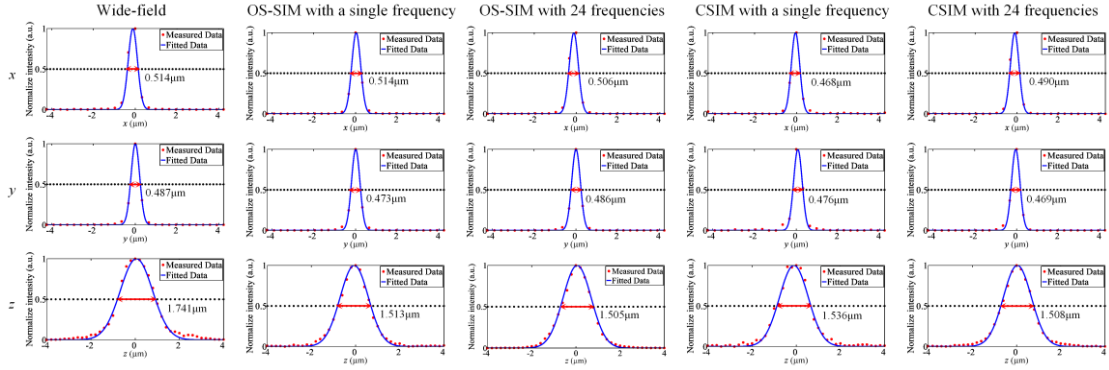

Fig. 8. Comparison of the imaging resolution of the wide-field illumination microscopy, OS-SIM with a single frequency, OS-SIM with 24 frequencies, CSIM with a single frequency, and CSIM with 24 frequencies.

The lateral resolution of the results reconstructed by OS-SIM and CSIM is not improved compared with the wide-field illumination microscopy. This is because the resolution of the reconstructed result is determined by the spatial frequency of the used structured light and the numerical aperture of the objective lens. Limited by our experimental setup, the spatial frequency of the used structured light ($0.331\ \mu m^{-1}$) was far lower than the cutoff frequency ($NA/\lambda \approx 1.596\ \mu m^{-1}$) of the objective lens, where $NA$ and $\lambda$ represent the numerical aperture of the objective lens and the wavelength of the illumination light, respectively.

The experimental results of optical-sectioning demonstrate that the proposed CSIM can reconstruct confocal images with higher SNR and has greater imaging depth than those of OS-SIM. This improvement is attributed to the integration of the confocal microscopy principle in CSIM, which allows for rejecting non-conjugate and defocus signals by extracting the conjugate signals to reconstruct confocal images.

The SNR of the reconstructed result is inversely proportional to the spatial frequency of the generated structured light. This is because the SNR of the reconstructed result is proportional to the contrast of the used structured light. Our experimental setup used an incoherent LED as the light source. The structured light generated by the LED has a strong DC background, contrasting the generated structured light decrease with the increase of the spatial frequency.

The proposed CSIM requires calibration of the conjugate relationship between the camera and the SLM. However, calibration does not affect the imaging process, as it is done before imaging. Calibration errors would make the extracted conjugate signals inaccurate, thus affecting the accuracy of the reconstructed confocal image. However, the results of our calibration experiments show that most of the calibrated reprojection error can be controlled within $\pm 0.6$ pixels. The experimental results show that the calibration error is tolerable.

As the camera and SLM often have different pixel sizes and counts, the calibrated coordinates of the points on the SLM's modulation plane conjugated to each camera pixel may be decimal numbers. Using these decimal numbers coordinates directly in Eq. (16) may lead to calculation errors. To minimize calculation errors, one can determine four integer pixels surrounding the decimal number coordinates and then use Eq. (16) to calculate the gray values of these integer pixels, and finally use bilinear interpolation to obtain the gray values of the decimal numbers based on the gray values of the integer pixels.

In conclusion, we have constructed the CSIM theoretical framework and developed its image reconstruction algorithm. The experimental results demonstrate that the proposed method yields a superior signal-to-noise ratio and greater imaging depth compared with existing OS-SIM. Our method has the potential to enhance the imaging performance of structured illumination microscopy and facilitate its wider application in various fields.

This work was financially supported by the Guangdong Basic and Applied Basic Research Foundation of China (2022A1515011560, 2023A1515011277) and the National Science Foundation of China (NSFC) (61875074).

**Author Contributions.**

**Jingang Zhong:** Conceptualization (equal); Formal analysis (equal); Funding acquisition (equal); Methodology (equal); Writing – original draft (equal); Writing – review & editing (equal). **Junzheng Peng:** Conceptualization (equal); Data curation (equal); Formal analysis (equal); Funding acquisition (equal); Investigation (equal); Methodology (equal); Software (equal); Validation (equal); Visualization (equal); Writing – original draft (equal); Writing – review & editing (equal). **Weishuai Zhou:** Data curation (equal); Form analysis (equal); Validation (equal); Visualization (equal). **Manhong Yao:** Validation (equal). **Xi Lin**: Visualization (equal). **Quan Yu**: Resource (equal).

**Author declarations.**

The authors declare no conflicts of interest.

## References

1. M. Minsky, "Microscopy Apparatus," United States Patent Office patent 3013467 (1957).
2. "Milestones in light microscopy," Nat Cell Biol **11**, 1165–1165 (2009).
3. A. Diaspro, G. Chirico, C. Usai, P. Ramoino, and J. Dobrucki, "Photobleaching," in *Handbook of Biological Confocal Microscopy: Third Edition* (2006), pp. 690–702.
4. M. A. A. Neil, R. Juškaitis, and T. Wilson, "Method of obtaining optical sectioning by using structured light in a conventional microscope," Opt. Lett. **22**, 1905–1907 (1997).
5. M. G. L. Gustafsson, "Surpassing the lateral resolution limit by a factor of two using structured illumination microscopy. SHORT COMMUNICATION," J Microsc **198**, 82–87 (2000).
6. R. Heintzmann and C. G. Cremer, "Laterally modulated excitation microscopy: improvement of resolution by using a diffraction grating," in *Proc.SPIE* (1999), Vol. 3568, pp. 185–196.
7. Z. Zhang, X. Ma, and J. Zhong, "Single-pixel imaging by means of Fourier spectrum acquisition," Nat. Commun. **6**, 6225 (2015).
8. J. Peng, M. Yao, J. Cheng, Z. Zhang, S. Li, G. Zheng, and J. Zhong, "Micro-tomography via single-pixel imaging," Opt. Express **26**, 31094–31105 (2018).
9. J. Peng, M. Yao, Z. Huang, and J. Zhong, "Fourier microscopy based on single-pixel imaging for multi-mode dynamic observations of samples," APL Photonics **6**, 046102 (2021).
10. J. Peng, M. Yao, Z. Cai, X. Qiu, Z. Zhang, S. Li, and J. Zhong, "Optical synthetic sampling imaging: Concept and an example of microscopy," Appl. Phys. Lett. **115**, 121101 (2019).
11. D. Lim, K. K. Chu, and J. Mertz, "Wide-field fluorescence sectioning with hybrid speckle and uniform-illumination microscopy," Opt. Lett. **33**, 1819 (2008).
12. D. Lim, T. N. Ford, K. K. Chu, and J. Mertz, "Optically sectioned in vivo imaging with speckle illumination HiLo microscopy," J. Biomed. Opt. **16**, 016014 (2011).

13. Z. Li, Q. Zhang, S.-W. Chou, Z. Newman, R. Turcotte, R. Natan, Q. Dai, E. Y. Isacoff, and N. Ji, "Fast widefield imaging of neuronal structure and function with optical sectioning in vivo," Sci. Adv. **6**, eaaz3870 (2020).
14. M. G. L. Gustafsson, L. Shao, P. M. Carlton, C. J. R. Wang, I. N. Golubovskaya, W. Z. Cande, D. A. Agard, and J. W. Sedat, "Three-Dimensional Resolution Doubling in Wide-Field Fluorescence Microscopy by Structured Illumination," Biophysical Journal **94**, 4957–4970 (2008).
15. Y. Wu and H. Shroff, "Faster, sharper, and deeper: structured illumination microscopy for biological imaging," Nat. Methods **15**, 1011–1019 (2018).
16. X. Chen, S. Zhong, Y. Hou, R. Cao, W. Wang, D. Li, Q. Dai, D. Kim, and P. Xi, "Superresolution structured illumination microscopy reconstruction algorithms: a review," Light Sci Appl **12**, 172 (2023).
17. M. Takeda and K. Mutoh, "Fourier transform profilometry for the automatic measurement of 3-D object shapes," Appl. Opt. **22**, 3977 (1983).
18. S. S. Gorthi and P. Rastogi, "Fringe projection techniques: Whither we are?," Opt Lasers Eng **48**, 133–140 (2010).
19. C. Zuo, T. Tao, S. Feng, L. Huang, A. Asundi, and Q. Chen, "Micro Fourier Transform Profilometry (μFTP): 3D shape measurement at 10,000 frames per second," Optics and Lasers in Engineering **102**, 70–91 (2018).
20. S. Gioux, A. Mazhar, and D. J. Cuccia, "Spatial frequency domain imaging in 2019: principles, applications, and perspectives," J. Biomed. Opt. **24**, 1 (2019).
21. N. Dögnitz and G. Wagnières, "Determination of tissue optical properties by steady-state spatial frequency-domain reflectometry," Laser Med Sci **13**, 55–65 (1998).
22. Z. Wang, T. Zhao, Y. Cai, J. Zhang, H. Hao, Y. Liang, S. Wang, Y. Sun, T. Chen, P. R. Bianco, K. Oh, and M. Lei, "Rapid, artifact-reduced, image reconstruction for super-resolution structured illumination microscopy," The Innovation **4**, 100425 (2023).
23. Z. Wang, T. Zhao, H. Hao, Y. Cai, K. Feng, X. Yun, Y. Liang, S. Wang, Y. Sun, P. R. Bianco, K. Oh, and M. Lei, "High-speed image reconstruction for optically sectioned, super-resolution structured illumination microscopy," Adv. Photon. **4**, (2022).
24. J. Jonkman and C. M. Brown, "Any Way You Slice It—A Comparison of Confocal Microscopy Techniques," J. Biomol. Tech. **26**, 54–65 (2015).
25. J. M. Murray, "Methods for Imaging Thick Specimens: Confocal Microscopy, Deconvolution, and Structured Illumination," Cold Spring Harb. Protoc. **2011**, 1399–1437 (2011).
26. P. Sen, B. Chen, G. Garg, S. R. Marschner, M. Horowitz, M. Levoy, and H. P. A. Lensch, "Dual photography," ACM Trans. Graph. **24**, 745–755 (2005).
27. Z. Zhang, X. Wang, G. Zheng, and J. Zhong, "Fast Fourier single-pixel imaging via binary illumination," Sci. Rep. **7**, 12029 (2017).
28. D. Li, W. Zhou, Z. Qiu, J. Peng, and J. Zhong, "Adaptive structured illumination optical-sectioning microscopy based on the prior knowledge of sample structure," Opt. Lasers. Eng. **172**, 107851 (2024).